\documentclass{aa}  

\usepackage{CJKutf8}
\newcommand{\zh}[1]{\begin{CJK*}{UTF8}{gbsn}#1\end{CJK*}}

\usepackage{graphicx}
\usepackage{txfonts}
\usepackage{hyperref}
\usepackage{natbib}
\usepackage{lscape}
\usepackage{pdflscape}
\usepackage{soul}

\hypersetup{
        colorlinks=true,    
        linkcolor=blue,  
        citecolor=blue,    
        filecolor=cyan,     
        urlcolor=black,      
} 

\begin{document} 

\title{Discovery of Two Hyperluminous Infrared Galaxies at $z\sim6$}

\titlerunning{Two HyLIRGs at $z\sim6$}
\authorrunning{Jin et al.}

\author{
Shuowen Jin (\zh{靳硕文})\inst{1,2,3,\thanks{Corresponding author: shuowen.jin@universite-paris-saclay.fr}},
Georgios E. Magdis\inst{2,3,4},
Raphael Gobat\inst{5},
Eva Schinnerer\inst{6},
Emanuele Daddi\inst{7},
Hervé Dole\inst{1},
Aurelien Henry\inst{8},
Daizhong Liu\inst{9},
Benjamin Magnelli\inst{7},
Francesca Rizzo\inst{10},
Nikolaj B. Sillassen\inst{2,3},
Sune Toft\inst{2,4},  
and Francesco Valentino\inst{2,3}
          }

   \institute{
   Université Paris-Saclay, CNRS, Institut d'Astrophysique Spatiale, 91405, Orsay, France
   \and     Cosmic Dawn Center (DAWN), Denmark
    \and
            DTU Space, Technical University of Denmark, Elektrovej 327, DK-2800 Kgs. Lyngby, Denmark
    \and
            Niels Bohr Institute, University of Copenhagen, Jagtvej 128, DK-2200 Copenhagen, Denmark
    \and
            Instituto de Física, Pontificia Universidad Católica de Valparaíso, Casilla 4059, Valparaíso, Chile
    \and
            Max-Planck-Institut für Astronomie, Königstuhl 17, 69117 Heidelberg, Germany
    \and
            Université Paris-Saclay, Université Paris Cité, CEA, CNRS, AIM, 91191 Gif-sur-Yvette, France
    \and
            Department of Physics, University of California, Merced, 5200 Lake Road, Merced, CA 92543, USA
    \and    
            Purple Mountain Observatory, Chinese Academy of Sciences, 10 Yuanhua Road, Nanjing 210023, China
    \and
            Kapteyn Astronomical Institute, University of Groningen, 9700 AV Groningen, The Netherlands
            }
   \date{Received XXX / Accepted XXX}

\abstract{
Hyperluminous infrared galaxies (HyLIRGs) at cosmic dawn are key to understanding the formation of the first massive galaxies. We report the spectroscopic confirmation of two unlensed HyLIRGs at $z\sim6$ in the COSMOS field, AzTECC63 and S2COS-84, originally selected from single-dish AzTEC and SCUBA-2 surveys. Using ALMA and NOEMA observations, we robustly detected CO and [C\,{\sc ii}] lines in both galaxies and unambiguously confirmed their redshifts to be $z=6.092$ and $z=5.90$, respectively. Panchromatic SED fitting reveals dust-obscured star formation rates of $\approx1000$ and $\approx2300\,M_\odot\,{\rm yr}^{-1}$, ranking them among the most intensely star-forming systems known at $z \gtrsim 6$. High-resolution ALMA observations reveal an extremely compact morphology for S2COS-84, with a dust-continuum effective radius of $\approx0.3$\,kpc, implying a star-formation rate surface density above the Eddington limit. The far-infrared SEDs of both galaxies, together with the compact size of S2COS-84, indicate warm dust ($T_{\rm dust} \approx 80$\,K) that is likely optically thick at far-infrared wavelengths. Both galaxies exhibit an extreme [C\,{\sc ii}] deficit, with $L_{\rm [C\,II]}/L_{\rm IR} \approx 10^{-3.7}$, an order of magnitude below that of normal star-forming galaxies. Their molecular gas masses, derived consistently from both [C\,{\sc ii}] and dust, together with their IR-derived SFRs imply short gas depletion times of $\approx40$--60\,Myr, making them plausible progenitors of the earliest massive quiescent galaxies found at $z \sim 5$.
}


\maketitle


\section{Introduction}

Hyperluminous infrared galaxies (HyLIRGs) are the most extreme star-forming systems in the Universe, defined by total infrared luminosities
(rest-frame 8--1000\,$\mu$m) of $L_{\rm IR} \geq 10^{13}\,L_\odot$, an order of magnitude above ultraluminous infrared galaxies (ULIRGs;
$L_{\rm IR} \geq 10^{12}\,L_\odot$; \citealt{Sanders1996ULIRG, Rowan-Robinson2000HyLIRG}).
These luminosities imply vigorous dust-obscured star formation rates (SFRs) above $1000~M_\odot~{\rm yr}^{-1}$ \citep{Kennicutt1998}. 
At such rates a galaxy can assemble $\sim$$10^{11}\,M_\odot$ of stars in only $\sim$100\,Myr.
The prototype, IRAS\,F10214+4724 at $z=2.29$, was identified in the IRAS survey \citep{RowanRobinson1991Natur}. Since then, wide-area far-infrared and (sub)millimeter surveys with \textit{Herschel}, the South Pole Telescope (SPT), and ground-based single-dish facilities have found HyLIRGs mostly at $z\sim1$--4 \citep[e.g.,][]{Ivison2019,Berman2022,Kamieneski2024,Liu_DZ2024NatAs}, forming the bright tail of the population of dusty star-forming galaxies \citep[DSFGs; see a review by][]{Casey2014DSFG-review}.
 
HyLIRGs play a central role in our understanding of how the most massive galaxies form. Their intense, short-lived, dust-enshrouded starbursts make them the leading candidate progenitors of the compact quiescent galaxies seen at $z\sim2$--3 and, eventually, of the massive elliptical galaxies that dominate the centers of present-day galaxy clusters \citep[e.g.,][]{Toft2014}. HyLIRGs often reside in overdense environments and frequently show signs of major mergers or host obscured active galactic nuclei \citep[AGN; e.g.,][]{Daddi2009GN20,Capak2011Nature,Engel2010merger,Rowan-Robinson2000HyLIRG,Ruiz2007HyLIRG}. This makes them natural laboratories for studying the joint growth of stellar bulges and supermassive black holes, and the early assembly of protoclusters \citep{Miller2018cluster_z4,Oteo2018cluster}.
JWST has made this connection more pressing by revealing a high abundance of massive quiescent galaxies at $z \gtrsim 3$--5 whose stellar populations imply formation epochs at $z\gtrsim6$--10 \citep[e.g.,][]{Carnall2023QGsample,Carnall2023z5,Glazebrook2024Natur,de_Graaff2025NatAs,Baker2025QG,Sillassen2026Vine}. Explaining these objects requires a population of extremely vigorous starbursts within the first billion years of cosmic history, which is precisely the regime in which HyLIRGs are expected to operate.

Advanced (sub)millimeter interferometers such as the Atacama Large Millimeter/submillimeter Array (ALMA) and the Northern Extended Millimeter Array (NOEMA) have spectroscopically confirmed distant DSFGs at $z>5$ and out to the epoch of reionization \citep[EoR, $z\sim7$; e.g.,][]{Walter2012, Vieira2010,Vieira2013,Riechers2013Nature, Riechers2017,Strandet2017,Zavala2017, Marrone2017Nature,Endsley2022z68,Fudamoto2021Nature,Hygate2023z7, Rowland2024z7}, proving these facilities to be powerful ``redshift machines'' \citep{Vieira2013,Neri2020noema,ChenCC2022alma,Cox2023zGal}.
Despite this, HyLIRGs at $z \gtrsim 5$ remain extremely rare. Only a handful of dusty starbursts with $L_{\rm IR} \gtrsim 10^{13}\,L_\odot$ have been confirmed at $z\gtrsim5$, and several of them are gravitationally lensed: for example, the $z \sim 5.3$--5.7 sources AzTEC-3, GN10, and CRLE \citep{Capak2011Nature, Riechers2020z5SFR,Pavesi2018}, the moderately lensed HFLS3 at $z = 6.34$ \citep{Riechers2013Nature}, and the strongly lensed SPT0311$-$58 at $z = 6.90$ \citep{Strandet2017,Marrone2017Nature}. Comparably luminous FIR emission is also seen in the host galaxies of some $z \sim 6$ quasars \citep[e.g.,][]{Bertoldi2003QSO, WangR2013QSO, Decarli2018QSOz6}, although the contribution of AGN heating is uncertain.
The known HyLIRGs at $z\sim6$ already strain theoretical expectations. Galaxy formation models struggle to reproduce the abundance of the brightest DSFGs only $\sim1$~Gyr after the Big Bang, and their large dust masses challenge models of dust production \citep[e.g.,][]{Hayward2013simu,Michalowski2015,Wilkins2018simu,Vijayan2022simu}. Each new $z\sim6$ HyLIRG therefore provides a valuable constraint on the formation pathways of the earliest massive galaxies.

In this Letter, we report the discovery of two unlensed HyLIRGs at $z\sim6$, selected from single-dish AzTEC and SCUBA-2 surveys and spectroscopically confirmed through ALMA and NOEMA detections of CO and [C\,{\sc ii}] lines.
We adopt a flat $\Lambda$CDM cosmology with $H_0=70$~km~s$^{-1}$~Mpc$^{-1}$ and $\Omega_{\rm M}=0.3$, and a Chabrier initial mass function \citep{Chabrier2003}.

\begin{figure*}[ht]
\centering
\includegraphics[width=0.95\textwidth]{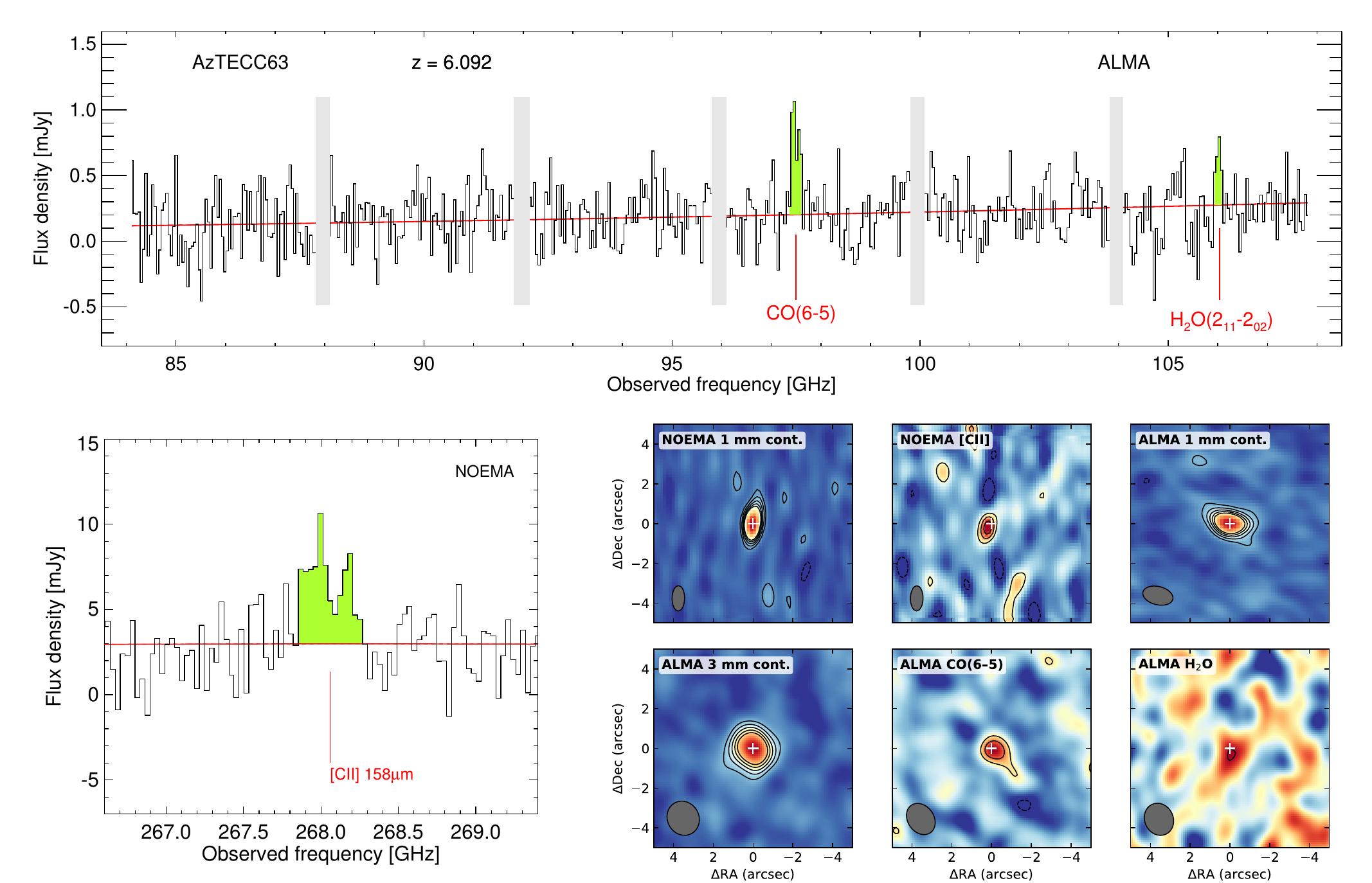}
\caption{ALMA and NOEMA spectra and images of AzTECC63 at $z=6.092$. The 3~mm spectrum is shown at the top and the 1~mm spectrum at the bottom left. Detected lines are highlighted in green and labeled. Gray areas mark the gaps between spectral windows. The red line shows the fit to the continuum. The bottom-right panels present the cleaned continuum and line images, with the beam shown in the corner. Contours start at $3\sigma$ and increase in steps of $2\sigma$.}
\label{fig1}
\end{figure*}

\begin{figure*}[ht]
\centering
\includegraphics[width=0.95\textwidth]{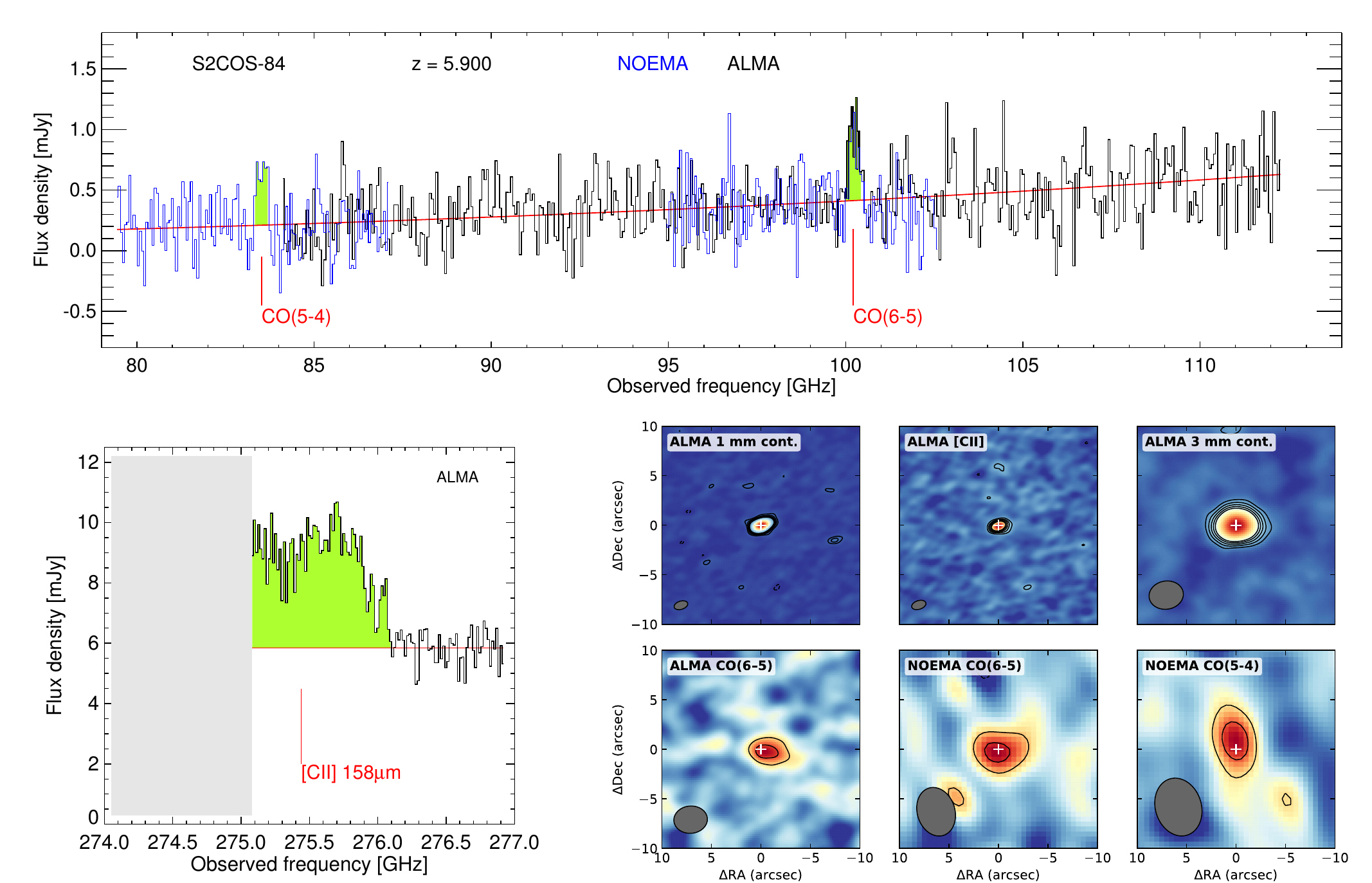}
\caption{Same as Fig.~\ref{fig1}, but for S2COS-84 at $z=5.90$. In bottom-left panel, the gray area marks the edge of ALMA Band~7. Only the low-resolution ALMA 1~mm continuum and [C\,{\sc ii}] data are shown here.}
\label{fig2}
\end{figure*}

\section{Selection and data}

\subsection{Selection}

The two galaxies, AzTECC63 and S2COS-84, were selected from submillimeter and millimeter surveys of the COSMOS field \citep{Scoville2007COSMOS}.
AzTECC63 was originally identified in the ASTE/AzTEC 1.1~mm survey by \cite{Aretxaga2011}.
It has no optical or near-infrared (NIR) counterpart (Fig.~\ref{fig:cutout}). Its high-resolution counterpart was identified in ALMA Band~6 as part of project 2013.1.00118.S (PI: M. Aravena), and was cataloged by \cite{Brisbin2017}.
S2COS-84 was first detected in the JCMT/SCUBA-2 850~$\mu$m survey \citep{Geach2017scuba2,Simpson2019SCUBA2} and is cataloged by \cite{Simpson2019SCUBA2} as S2COS850.0084. Its high-resolution counterpart was revealed by the ALMA Band~7 program 2016.1.00463.S (PI: Y. Matsuda).
Benefiting from the rich multiwavelength data in COSMOS, we estimated their redshifts to be $z\gtrsim6$ based on their FIR SEDs and archival ALMA spectroscopy. As candidate $z>6$ DSFGs, we then observed them with ALMA and NOEMA to securely confirm their redshifts via line detections.

\subsection{ALMA and NOEMA observations}

AzTECC63 was observed with ALMA 3~mm line scans as part of our program 2022.1.00884.S (PI: R. Gobat). Following the line detections with ALMA (Fig.~\ref{fig1}), we carried out NOEMA 1~mm follow-up observations of the [C\,{\sc ii}] 158~$\mu$m line (ID: S23CT; PIs: S. Jin, A. Henry).
S2COS-84 was observed with ALMA 3~mm line scans as part of program 2021.1.00246.S (PI: C. Chen), in which a single line was detected at 100.2~GHz with a width of $\sim1100$~km~s$^{-1}$ (Fig.~\ref{fig2}). To robustly confirm its redshift, we followed it up with NOEMA to detect additional CO lines (ID: S24CL; PIs: S. Jin, E. Schinnerer) and with ALMA to detect [C\,{\sc ii}] (ID: 2024.1.00945.S; PI: R. Gobat). The ALMA [C\,{\sc ii}] observations of S2COS-84 were carried out in both compact (C3) and extended (C6) configurations in Band~7, yielding angular resolutions of $0.91''$ and $0.12''$, respectively. S2COS-84 is unresolved in the low-resolution data and resolved in the high-resolution ones. We use the low-resolution data for the redshift confirmation, and the combination of low- and high-resolution data for the size measurements.
 
The raw data from the ALMA programs listed above were reduced and calibrated using the standard ALMA pipeline in CASA \citep{McMullin2007CASA}. Following our established procedure \citep{Jin2019alma,Jin2022,Jin2024lens}, we converted the calibrated measurement sets to \texttt{uvfits} format for further analysis in the $uv$ plane with the GILDAS software. The NOEMA observations were reduced and calibrated at IRAM Grenoble with the GILDAS pipeline, producing calibrated $uv$ tables. Both the ALMA and NOEMA data were then analyzed in the $uv$ plane using GILDAS. The 1D spectra were extracted by running the GILDAS \texttt{uvfit} routine on the $uv$ tables at all frequencies, adopting a point-source model at the fixed position of the ALMA continuum peak. The continuum and line images were produced using natural weighting and the GILDAS H\"ogbom clean algorithm.
For ALMA 1~mm spectrum of S2COS-84, we performed two versions of spectrum extractions: (1) \texttt{uvfit} on the low-resolution $uv$ tables using point-source model; (2) an aperture extraction in the clean data cube generated with CASA \texttt{tclean} by combining high- and low-resolution data. The two versions of spectra agree well, giving consistent continuum and [C\,{\sc ii}] fluxes.

\subsection{Ancillary data}

AzTECC63 is undetected in the optical, NIR, and IRAC images. S2COS-84 is well detected in the NIR and has photometry in the COSMOS2020 catalog \citep{Weaver2022COSMOS2020}; we therefore adopt its NIR and IRAC photometry for the SED fitting. 
To measure their FIR-to-radio photometry, we applied our super-deblending technique \citep{Liu_DZ2017,Jin2018cosmos}. We followed the procedure of \cite{Jin2018cosmos} with a more complete prior catalog of $\sim$200k sources drawn from the COSMOS2020 \citep{Weaver2022COSMOS2020}, radio \citep{Smolcic2017,van_der_Vlugt2021COSMOS-XS}, and ALMA \citep{Liu2019A3COSMOS} catalogs. In addition, we updated the SCUBA-2 and radio photometry by deblending the latest SCUBA-2 \citep{Simpson2019SCUBA2} and MeerKAT \citep{Hale2025MeerKAT} images.
The archival ALMA continuum fluxes of the two galaxies are available in the A3COSMOS catalog \citep{Liu2019A3COSMOS}, and we adopt them directly.
Neither galaxy is covered by the JWST COSMOS-Web survey \citep{Casey2023cosmosweb}.

Neither source shows evidence of gravitational lensing, as no foreground galaxies with $\log M_\star/M_\odot > 10.5$ are found within 15\arcsec\ of either source in the COSMOS2020 catalog. No X-ray galaxy groups from \citet{Gozaliasl2019Xray} lie within 2\arcmin\ of either source; the nearest groups ($M_{200} \approx 0.5, 2.3 \times 10^{14}\,M_\odot$ at $z = 0.4$, 1.0) are located at 1.8 and 3\,$R_{200}$. Adopting a singular isothermal sphere lens model \citep{Narayan1996lensing}, this implies a magnification of $\mu < 1.04$ for each.

\section{Results}
\subsection{Redshift confirmation}

We present the ALMA and NOEMA spectra of the two galaxies in Figs.~\ref{fig1} and \ref{fig2}; two to three lines are detected in each galaxy. For AzTECC63, the CO(6--5) and [C\,{\sc ii}] lines are detected at $>6\sigma$, yielding a robust spectroscopic redshift of $z=6.092$. The images in Fig.~\ref{fig1} show well-matched peak positions for the continuum and the lines, and the line frequencies are fully consistent with this redshift. The ${\rm H_2O(2_{11}}$--${\rm 2_{02})}$ line is also tentatively detected at $3\sigma$ at the expected frequency.
For S2COS-84, three lines are detected at $z=5.90$: CO(5--4) with NOEMA, CO(6--5) with both ALMA and NOEMA, and [C\,{\sc ii}] with ALMA (Fig.~\ref{fig2}). We measured the line fluxes from the spectra and list them in Table~\ref{tab:physpars}. We note that the [C\,{\sc ii}] line of S2COS-84 falls at the edge of ALMA Band~7 and its red wing is partly missing; the redshift is therefore derived from the two CO lines. Its [C\,{\sc ii}] flux is obtained by integrating over the observed [C\,{\sc ii}] channels and adding an estimate of the missing flux, derived by assuming the CO(5--4) line profile, with the corresponding uncertainty propagated.

\subsection{Size measurements of S2COS-84}
 
AzTECC63 was only observed in compact configurations and is unresolved. S2COS-84 was observed in both compact and extended configurations in ALMA Band~7 and is resolved in the high-resolution data.
We measured the dust continuum size of S2COS-84 in the $uv$ plane,
which avoids the deconvolution and correlated-noise biases that affect image-plane fits to marginally resolved sources \citep[e.g.,][]{Tan2024a,Tan2024Natur}. We combined the continuum visibilities of the line-free channels.
Following \cite{Tan2024a}, we modeled the emission with an elliptical Spergel profile \citep{Spergel2010profile}, which has an analytic Fourier transform, making it well suited to interferometric data, and closely approximates a S\'ersic profile \citep{Tan2024a}.
The best fit (Fig.~\ref{fig:uvfit}) gives $R_{\rm e}=0.052\pm0.002\arcsec$ ($0.31\pm0.01$\,kpc; circularized $0.047\arcsec$, i.e., 0.28\,kpc), an axis ratio of $q=0.79\pm0.02$, a position angle of $56\pm3^\circ$, and $\nu\approx0.0$--$0.2$. Using the calibration of \citet{Tan2024a}, this corresponds to a pseudo-S\'ersic index of $n=1.3\pm0.2$,
close to an exponential disk. Fitting S\'ersic profiles directly to the
visibilities gives consistent results ($n=1.1$--$1.3$ and the same $R_{\rm e}$).
Both a Gaussian ($n=0.5$) and a de~Vaucouleurs-like profile ($n=4$) are strongly disfavored ($\Delta\chi^2>100$).
 
Applying the same procedure to the continuum-subtracted [C\,{\sc ii}] visibilities yields $R_{\rm e,[CII]}=0.138\pm0.012\arcsec$ ($0.82\pm0.07$\,kpc; circularized $0.128\arcsec$, i.e., 0.75\,kpc), $q=0.85\pm0.11$, and $\nu=0.6\pm0.7$,
corresponding to a pseudo-S\'ersic index of $n\approx1$, consistent with an exponential disk. These results reveal a compact disk-like morphology for S2COS-84 in both the dust continuum and the [C\,{\sc ii}] emission.
The [C\,{\sc ii}] emission is $2.7\pm0.3$ times more extended than the
dust continuum. [C\,{\sc ii}] is commonly found to be more extended than the dust in $z\gtrsim4$ star-forming galaxies, typically by factors of
$\sim$1.5--2 \citep{Gullberg2018CIIhalo,Fujimoto2019CIIhalo,Ikeda2025CII}; S2COS-84 lies at the extended end of this distribution.

\subsection{SFR, stellar mass, and dust properties}
\label{sec:dust}
 
We performed panchromatic SED fitting for the two sources with the MICHI2 code \citep{Liu2021michi2}, using their NIR, IRAC, super-deblended FIR-to-radio, and ALMA/NOEMA continuum photometry. The best-fit SEDs are shown in Fig.~\ref{fig:sed}; each is fitted with five components: stellar \citep{BC03}, mid-infrared AGN \citep{Mullaney2011}, warm and cold dust \citep{Draine2007SED}, and radio \citep{Magnelli2015}. The derived physical parameters are listed in Table~\ref{tab:physpars}. The best fits give infrared luminosities of $\log(L_{\rm IR}/L_\odot)=13.02\pm0.03$ and $13.36\pm0.04$ for AzTECC63 and S2COS-84, respectively, confirming their HyLIRG nature. These luminosities correspond to ${\rm SFR_{IR}}\approx1000$ and $2300~M_\odot~{\rm yr}^{-1}$.
As AzTECC63 is undetected in IRAC and at shorter wavelengths, we can only place an upper limit on its stellar mass, $M_*<4.6\times10^{10}~M_\odot$. The stellar mass of S2COS-84 is constrained to $M_*\approx10^{11}~M_\odot$. These stellar masses and SFRs place both galaxies well above the star-forming main sequence at $z\sim6$ \citep{Cole2025MS}, confirming their starburst nature. Their radio fluxes are well reproduced by the typical infrared--radio correlation \citep{Magnelli2015}, and thus show no evidence of strong radio AGN activity.

Compared with the literature, the two galaxies are among the most intensely star-forming systems known at $z\gtrsim6$. Their SFRs are 3--13 times higher than those of MAMBO-9 at $z=5.85$ \citep{Jin2019alma,Casey2019}, the lensed G09~83808 at $z=6.02$ \citep{Zavala2017}, and REBELS-25 at $z=7.31$ \citep{Hygate2023z7,Rowland2024z7}. The only $z\gtrsim6$ systems with comparable SFRs are the lensed SPT0311$-$58 at $z=6.9$ \citep{Strandet2017,Marrone2017Nature} and the dust-obscured quasars COS-87259 at $z=6.8$ \citep{Endsley2022z68} and GNz7q at $z=7.19$ \citep{Fujimoto2022Nature}, which are among the most extreme objects known to date.

To constrain the dust properties, we fit the FIR SEDs of both galaxies with optically thin and general-opacity modified blackbody models, including the effect of the CMB \citep{daCunha2013}. We adopt a dust mass absorption coefficient of $\kappa_\nu=\kappa_0(\nu/\nu_0)^\beta$ with $\kappa_0=0.43~{\rm cm^2~g^{-1}}$ at 850~$\mu$m \citep{Li-Draine2001dust}. 
The best-fit SEDs and parameters are presented in Fig.~\ref{fig:FIRsed}. For AzTECC63, the general-opacity model is mildly preferred ($\sim2.3\sigma$), suggesting optically thick dust in the FIR. For S2COS-84, the two models are statistically indistinguishable. However, two pieces of evidence disfavor the optically thin case: (1) the thin fits require unusually low emissivity indices, $\beta\approx1.1$--1.25, for both galaxies; and (2) for S2COS-84, the dust mass from the thin fit, distributed over its compact size of $R_{\rm e}=0.28$~kpc, implies a high optical depth of $\tau_{158}\approx2.6$ at rest-frame 158~$\mu$m, which is inconsistent with the optically thin assumption. We therefore conclude that the dust in both galaxies is likely optically thick in the FIR, with warm dust temperatures of $T_{\rm dust}\approx80$~K, although high-resolution observations are required to measure the size of AzTECC63 and confirm this.

\subsection{$L_{\rm [CII]}$--$L_{\rm IR}$ relation, gas mass and depletion time}
 
\begin{figure}
\centering
\includegraphics[width=0.49\textwidth]{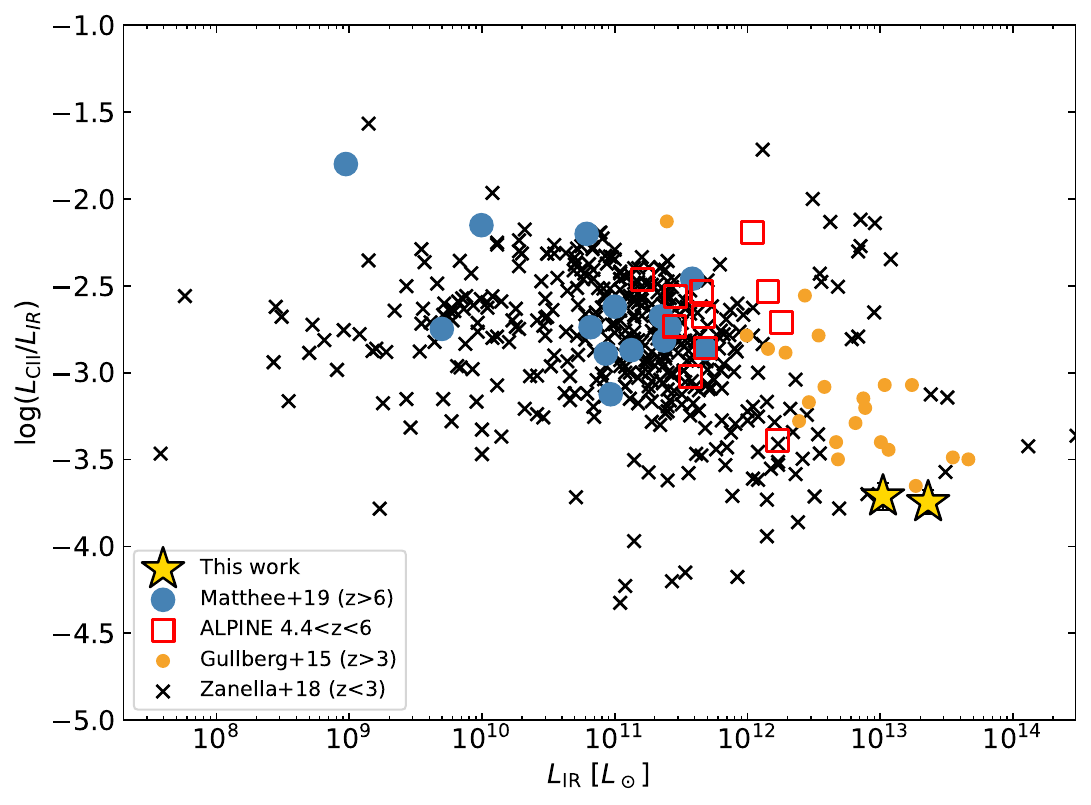}
\caption{
$L_{\rm [CII]}/L_{\rm IR}$ versus $L_{\rm IR}$ for the two galaxies of this work and for literature samples. The $z>6$ sources are from \cite{Matthee2019CII}, with $L_{\rm IR}$ estimated by \cite{Schaerer2020Cii} assuming $T_{\rm dust}=45$~K. The ALPINE sample is from \cite{Schaerer2020Cii}. The $z>3$ and $z<3$ samples are from \cite{Gullberg2015Cii} and \cite{Zanella2018Cii}, respectively.
}
\label{fig-CII-IR}
\end{figure}

Figure~\ref{fig-CII-IR} presents $L_{\rm [CII]}/L_{\rm IR}$ as a function of $L_{\rm IR}$ for the two galaxies and for literature samples \citep{Matthee2019CII,Schaerer2020Cii,Gullberg2015Cii,Zanella2018Cii}. Both galaxies have $L_{\rm [CII]}/L_{\rm IR}\approx10^{-3.7}$. These ratios are comparable to those of local ULIRGs \citep{Diaz-Santos2013CII} and about one order of magnitude lower than those of normal star-forming galaxies, $L_{\rm [CII]}/L_{\rm IR}\sim10^{-2.5}$
\citep{Malhotra2001ISM,Diaz-Santos2013CII,Smith2017CII,Herrera-Camus2018CII}.

From their [C\,{\sc ii}] luminosities, we estimate molecular gas masses of
$M_{\rm mol} \approx 6.3 \times 10^{10}$ and $1.3 \times 10^{11}\,M_\odot$ for AzTECC63 and S2COS-84, respectively, adopting the [C\,{\sc ii}]-to-H$_2$ conversion factor $\alpha_{\rm [C\,II]} = 31\,M_\odot/L_\odot$ of \citet{Zanella2018Cii}. These are in excellent agreement with the gas masses inferred from the dust masses of our optically thick SED fits, $\log(M_{\rm mol}/M_\odot) = 10.66 \pm 0.17$ and $11.10 \pm 0.12$, assuming a typical gas-to-dust ratio of $\delta_{\rm GDR} = 100$ \citep[e.g.,][]{Magdis2012SED}. This suggests that, despite their strong [C\,{\sc ii}] deficits, [C\,{\sc ii}] remains a reliable tracer of the molecular gas in these galaxies. Combined with their IR-derived SFRs, these gas masses imply short depletion times of $\tau_{\rm dep} \approx 40$--60\,Myr. In the absence of significant gas replenishment, their cold gas reservoirs would thus be exhausted by $z \approx 5.6$--5.8, quenching their star formation well before $z \sim 5$. This makes them plausible progenitors of the earliest massive quiescent galaxies identified at $z \gtrsim 4$--5 \citep[e.g.,][]{Carnall2023z5, de_Graaff2025NatAs}.

\begin{table}
    \caption{Physical properties}
    \centering
    \setlength{\tabcolsep}{2pt}
    \renewcommand\arraystretch{1.5}
    \begin{tabular}{c c c c c c c c}
    \hline\hline
        ID  & AzTECC63 & S2COS-84 \\
        RA  & 149.83755 & 149.58108\\
        Dec & 2.51922 & 2.25228\\
        $z$ & $6.092\pm0.001$ & $5.900\pm0.002$\\
        \hline
        CO(6--5) [mJy~km~s$^{-1}$] & $432\pm72$ &  $870\pm106$\\
        CO(5--4) [mJy~km~s$^{-1}$] & -- &  $710\pm175$\\
        FWHM [km~s$^{-1}$] & $760\pm205$ & $1100\pm140$ \\
        \hline
        ${\rm H_2O(2_{11}}$--${\rm 2_{02})}$ [mJy~km~s$^{-1}$]  & $197\pm59$ &  $<450$ \\
        \hline
        [C\,{\sc ii}] [Jy~km~s$^{-1}$] & $2.11\pm0.34$ &  $4.48\pm0.59$\\
        $L_{\rm [CII]}$ [$10^9~L_{\odot}$] & $2.03\pm0.33$ &  $4.11\pm0.54$\\
         \hline
        $R_{\rm e,cont}$ [kpc] &  -- & $0.28\pm0.01$\\
        $R_{\rm e,[CII]}$ [kpc] &  -- & $0.75\pm0.07$\\
        \hline
        log($M_\ast/M_\odot$) & $<10.65$ & $10.91\pm0.25$ \\
        log($L_{\rm IR}/L_\odot$) & $13.02\pm0.03$ & $13.36\pm0.04$\\
        \hline
        log($M_{\rm dust,thin}/M_\odot$) & $9.13\pm0.10$ & $9.33\pm0.04$\\ 
        $T_{\rm dust,thin}$ [K] & $48^{+6}_{-7}$ & $57^{+4}_{-4}$\\
        \hline
        log($M_{\rm dust,thick}/M_\odot$) & $8.66\pm0.17$ & $9.10\pm0.12$\\ 
        $T_{\rm dust,thick}$ [K] & $80^{+10}_{-12}$ & $80^{+11}_{-12}$\\
\hline\hline
    \end{tabular}
    \label{tab:physpars}
\end{table}

\section{Discussion}

The [C\,{\sc ii}] deficits of the two galaxies are among the most extreme found in DSFGs at $z>3$. The compact morphology of S2COS-84 ($R_{\rm e}\approx0.3$~kpc) is rarely seen in dusty galaxies at $z>1$, but is comparable to that of local ULIRGs \citep{Pereira-Santaella2020}. We suspect that these two extreme properties are causally connected, with the compact morphology contributing to the [C\,{\sc ii}] deficit. The small size and high SFR of S2COS-84 imply an extremely high SFR surface density, $\Sigma_{\rm SFR}={\rm SFR}/(2\pi R_{\rm e}^2)\approx4700~M_\odot~{\rm
yr^{-1}~kpc^{-2}}$. This is more than two orders of magnitude above the values of DSFG samples in the literature \citep{Hodge2019,Simpson2019SCUBA2,Jin2022} and exceeds the radiation-pressure (Eddington) limit of $\sim10^{3}~M_\odot~{\rm yr^{-1}~kpc^{-2}}$ expected for dusty starbursts \citep{Thompson2005}.

Such a compact, dust-rich configuration implies a significant dust opacity at the wavelength of [C\,{\sc ii}]. The general-opacity fit to the FIR SED of S2COS-84 (Sect.~\ref{sec:dust}) yields $\lambda_0=168^{+58}_{-86}~\mu$m and $T_{\rm dust}=80^{+11}_{-12}$~K, that is, $\tau_{158}\approx1.1$; fixing the emitting area to $2\pi R_{\rm e}^2$ instead gives $T_{\rm dust}=104\pm3$~K and $\tau_{158}\approx2.7$. For gas that is well mixed with the dust, the [C\,{\sc ii}] line is then attenuated by only 0.2--0.5~dex, leaving $L_{\rm [CII]}/L_{\rm IR}\approx10^{-3.5}$--$10^{-3.2}$ after correction, still far below the $\sim10^{-2.5}$ of normal star-forming galaxies. Moreover, since the [C\,{\sc ii}] emission is 2.7 times more extended than the dust continuum, part of the line emission arises outside the optically thick core, so this attenuation should be regarded as an upper limit. Only an unrealistic foreground-screen geometry with $\tau_{158}\approx2.7$ could explain the full deficit. Dust opacity thus contributes to, but cannot fully explain, the [C\,{\sc ii}] deficit. The remainder is likely driven by the extreme conditions of the compact starburst: thermal saturation of the [C\,{\sc ii}] upper level at high gas temperature and density \citep{Munoz2016CII}, reduced photoelectric heating efficiency on charged grains \citep{Malhotra2001ISM}, and absorption of ionizing photons by dust within H\,{\sc ii} regions \citep{Abel2009ULIRG}.

Alternatively, part of $L_{\rm IR}$ could be powered by a dust-obscured AGN \citep[e.g.,][]{McKinney2021AGN}, which would lower the SFR and $\Sigma_{\rm SFR}$ and raise the intrinsic $L_{\rm [CII]}/L_{\rm IR}$ of the star-forming component. The super-Eddington $\Sigma_{\rm SFR}$ and the hot, compact dust are consistent with such a central heating source: an optically thick dust photosphere radiating $L_{\rm IR}\approx2\times10^{13}~L_\odot$ at $T_{\rm dust}\approx100$~K has a radius of only $\sim0.4$~kpc, comparable to the measured size. However, this holds for any central power source and does not by itself require an AGN. Moreover, the dust continuum is resolved with a disk-like profile ($n\approx1.3$, $q\approx0.8$) rather than a point source, which argues against a dominant unresolved nucleus at 158~$\mu$m, although a large optical depth would hide the central source at this wavelength. The radio emission shows no excess with respect to the infrared--radio correlation, which disfavors a radio-loud AGN but does not exclude a radio-quiet one. Finally, an AGN would need to contribute $\gtrsim80\%$ of $L_{\rm IR}$ to account for the remaining deficit or to bring $\Sigma_{\rm SFR}$ below the Eddington limit, whereas a modest contribution (20--50\%) changes these quantities by only 0.1--0.3~dex. Rest-frame mid-infrared observations with JWST are required to constrain the AGN fraction.

\section{Conclusions}

Using ALMA and NOEMA observations, we have spectroscopically confirmed two unlensed HyLIRGs, AzTECC63 at $z = 6.092$ and S2COS-84 at $z = 5.90$. With dust-obscured SFRs of $\approx1000$ and $2300\,M_\odot\,{\rm yr}^{-1}$, they rank among the most intensely star-forming systems known at $z \gtrsim 6$. Both galaxies exhibit an extreme [C\,{\sc ii}] deficit, $L_{\rm [C\,II]}/L_{\rm IR} \approx 10^{-3.7}$, yet their [C\,{\sc ii}]- and dust-based molecular gas masses agree. Combined with their SFRs, these imply short gas depletion times of $\approx40$--60\,Myr, making them plausible progenitors of the earliest massive quiescent galaxies found at $z \sim 5$.
High-resolution ALMA imaging shows that S2COS-84 is extremely compact, with a dust-continuum effective radius of $R_{\rm e} = 0.28$\,kpc and a [C\,{\sc ii}] extent 2.7 times larger. The implied $\Sigma_{\rm SFR} \approx 4700\,M_\odot\,{\rm yr}^{-1}\,{\rm kpc}^{-2}$ exceeds the Eddington limit for dusty starbursts. The FIR SEDs, together with this compact size, indicate warm dust ($T_{\rm dust} \approx 80$ \,K) that is likely optically thick in the FIR. However, in S2COS-84, dust opacity can only partially account for the [C \,{\sc ii}] deficit, which further requires either the extreme conditions of the compact starburst or a dominant contribution to $L_{\rm IR}$ from a dust-obscured AGN. High-resolution imaging of AzTECC63 and rest-frame mid-infrared observations of both galaxies with JWST are needed to establish what powers these extreme systems.

\begin{acknowledgements}
This work is based on observations carried out under projects number S23CT and S24CL with the IRAM Interferometer NOEMA. IRAM is supported by INSU/CNRS (France), MPG (Germany) and IGN (Spain). We thank the IRAM staff for help provided during the observations and for data reduction.
This paper makes use of the following ALMA data: ADS/JAO.ALMA\#2024.1.00945.S, 2024.1.00467.S, 2022.1.00884.S, 2021.1.00246.S, 2021.1.00705.S, 2016.1.00463.S, and 2013.1.00118.S. ALMA is a partnership of ESO (representing its member states), NSF (USA) and NINS (Japan), together with NRC (Canada), NSTC and ASIAA (Taiwan), and KASI (Republic of Korea), in cooperation with the Republic of Chile. The Joint ALMA Observatory is operated by ESO, AUI/NRAO and NAOJ.
SJ acknowledges the UPSaclay STAR $\phi$ program (Grant Agreement ID: 101216532) from the European Union’s COFUND action, a part of the Marie Sk\l{}odowska-Curie Actions Program within the European Commission MSCA framework. Views and opinions expressed are however those of the author(s) only and do not necessarily reflect those of the European Union or European Research Executive Agency. Neither the European Union nor the REA can be held responsible for them.
GEM and SJ acknowledge the Villum Fonden research grants 37440 and 13160.
DL acknowledges the support from the Strategic Priority Research Program of the Chinese Academy of Sciences, grant No. XDB0800401.
The Cosmic Dawn Center (DAWN) is funded by the Danish National Research Foundation under grant DNRF140.

\end{acknowledgements}

\bibliographystyle{aa}
\bibliography{biblio}

\begin{appendix}

\section{Supporting material}

\begin{figure}%
\centering
\includegraphics[width=0.4\textwidth]{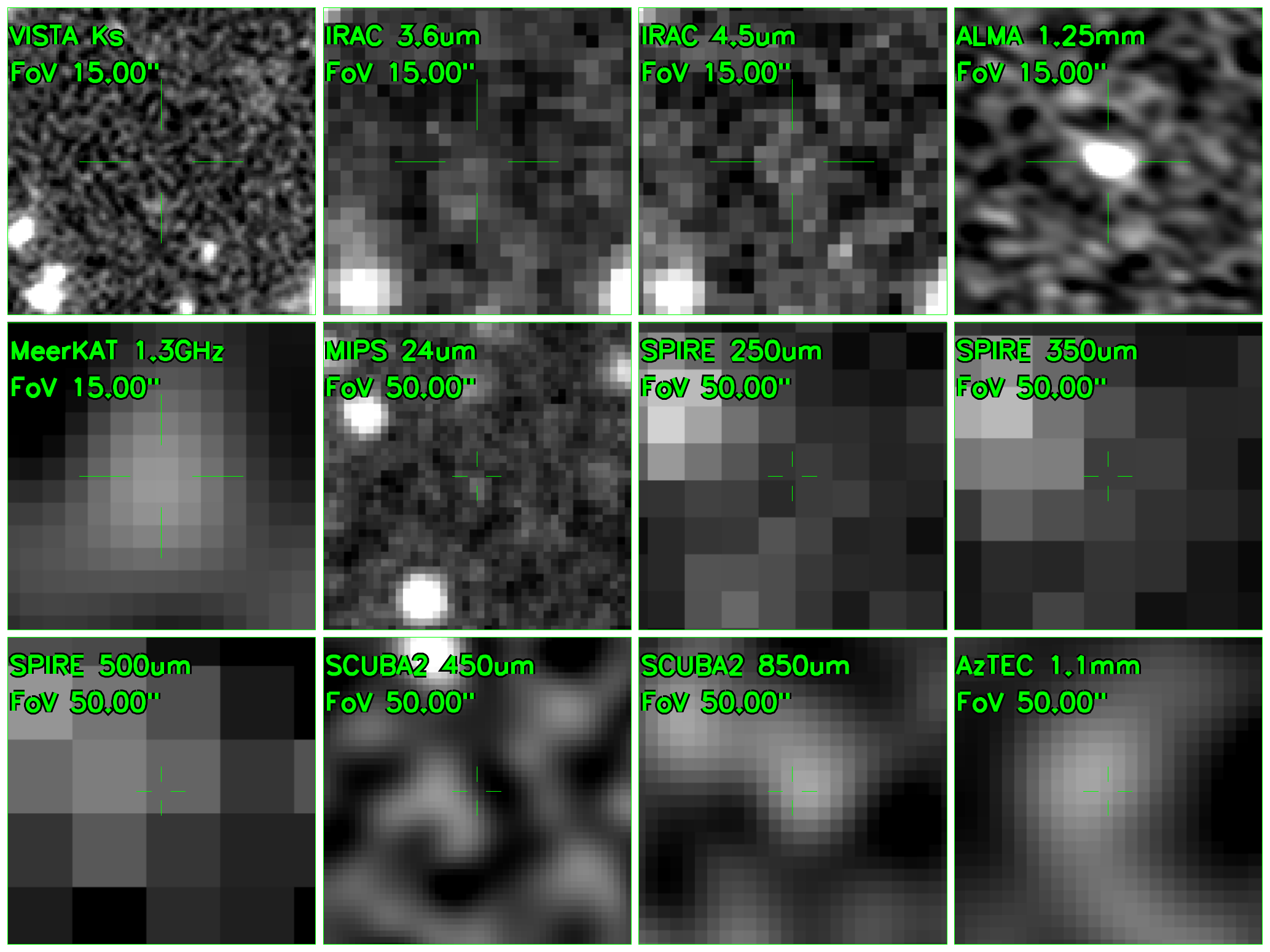}
\includegraphics[width=0.4\textwidth]{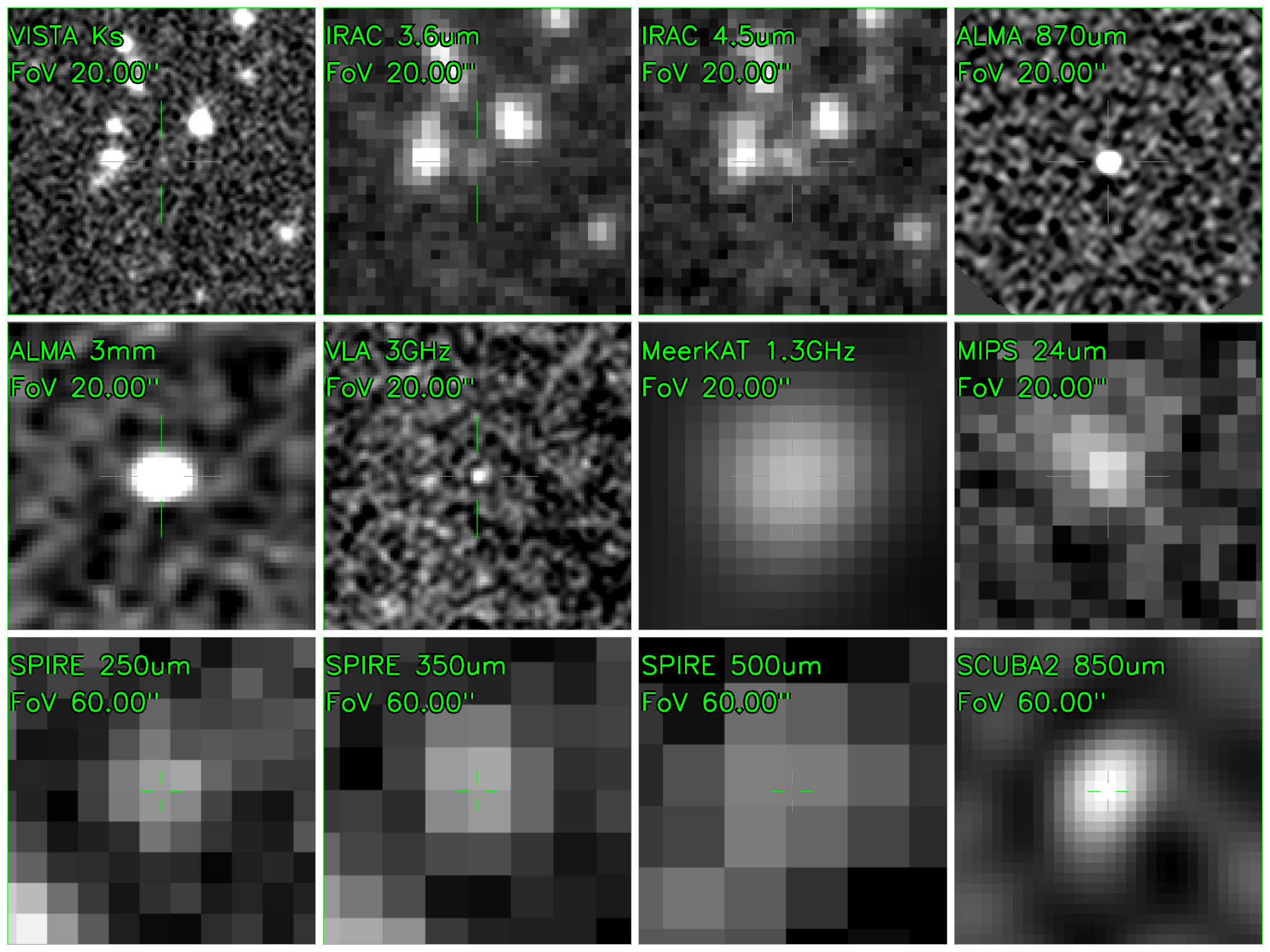}
\caption{Multi-wavelength images of the two galaxies, AzTECC63 (top) and S2COS-84 (bottom). The dataset and FoV are labeled in green text.
}
\label{fig:cutout}
\end{figure}

\begin{figure}%
\centering
\includegraphics[width=0.49\textwidth]{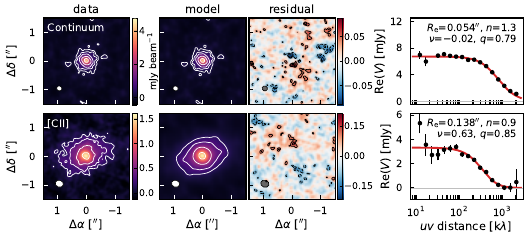}
\caption{Dust continuum and [C\,{\sc ii}] images of S2COS-84 and the corresponding $uv$-plane fits. The data and model panels show dirty images (natural weighting) of the observed and best-fit model visibilities. Contours start at $10\sigma$ for the continuum and at $3\sigma$ for [C\,{\sc ii}], and increase by factors of 2. Residual contours are drawn at $\pm2$, $\pm3$, and $\pm4\sigma$. The right panels show the Spergel profile fits to the visibilities.
}
\label{fig:uvfit}
\end{figure}

\begin{figure}%
\centering
\includegraphics[width=0.45\textwidth]{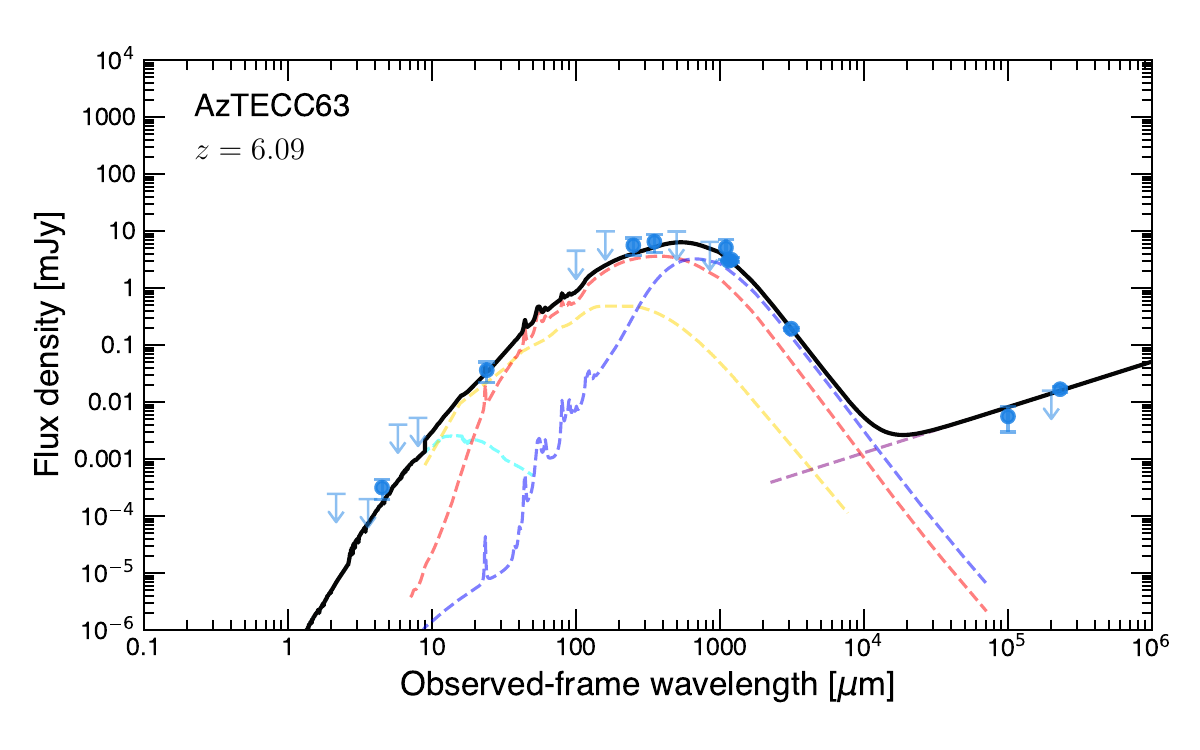}
\includegraphics[width=0.45\textwidth]{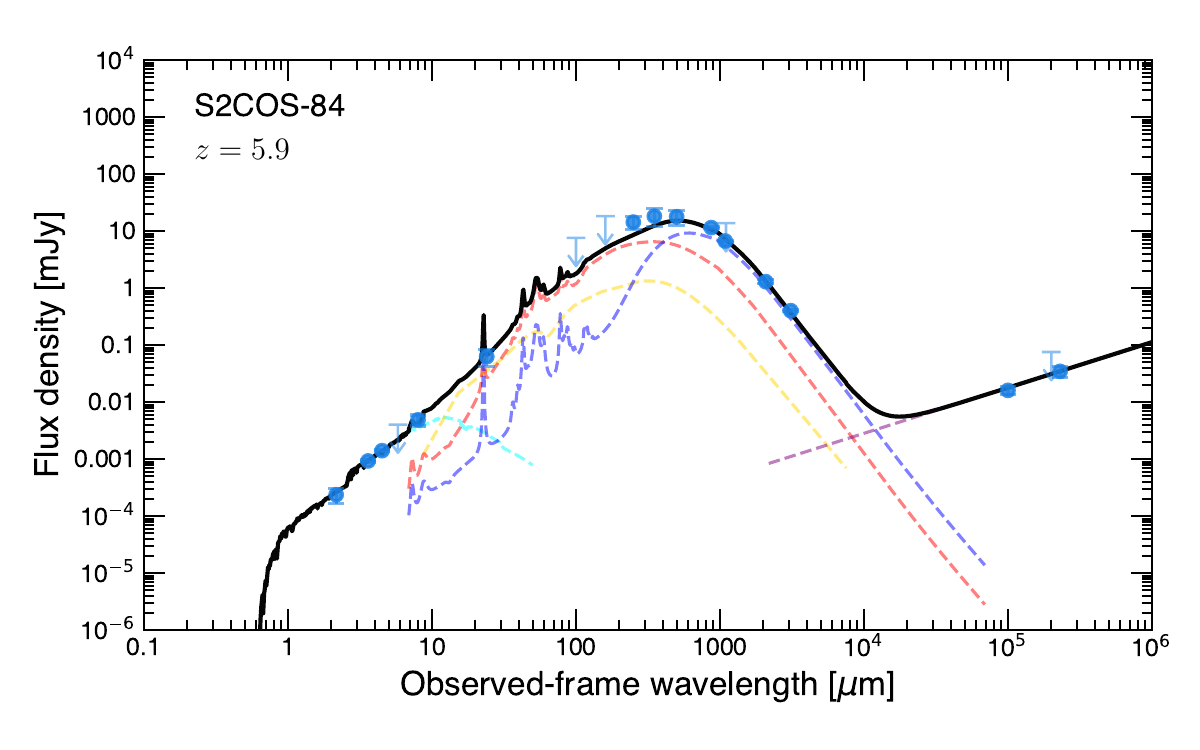}
\caption{Panchromatic SEDs of the two galaxies fitted with the MICHI2 code \citep{Liu2021michi2}.
The black curve shows the best fit, which is the sum of five components: stellar (cyan), mid-infrared AGN (yellow), warm dust (red), cold dust (blue), and radio (purple).
}
\label{fig:sed}
\end{figure}

\begin{figure}%
\centering
\includegraphics[width=0.43\textwidth]{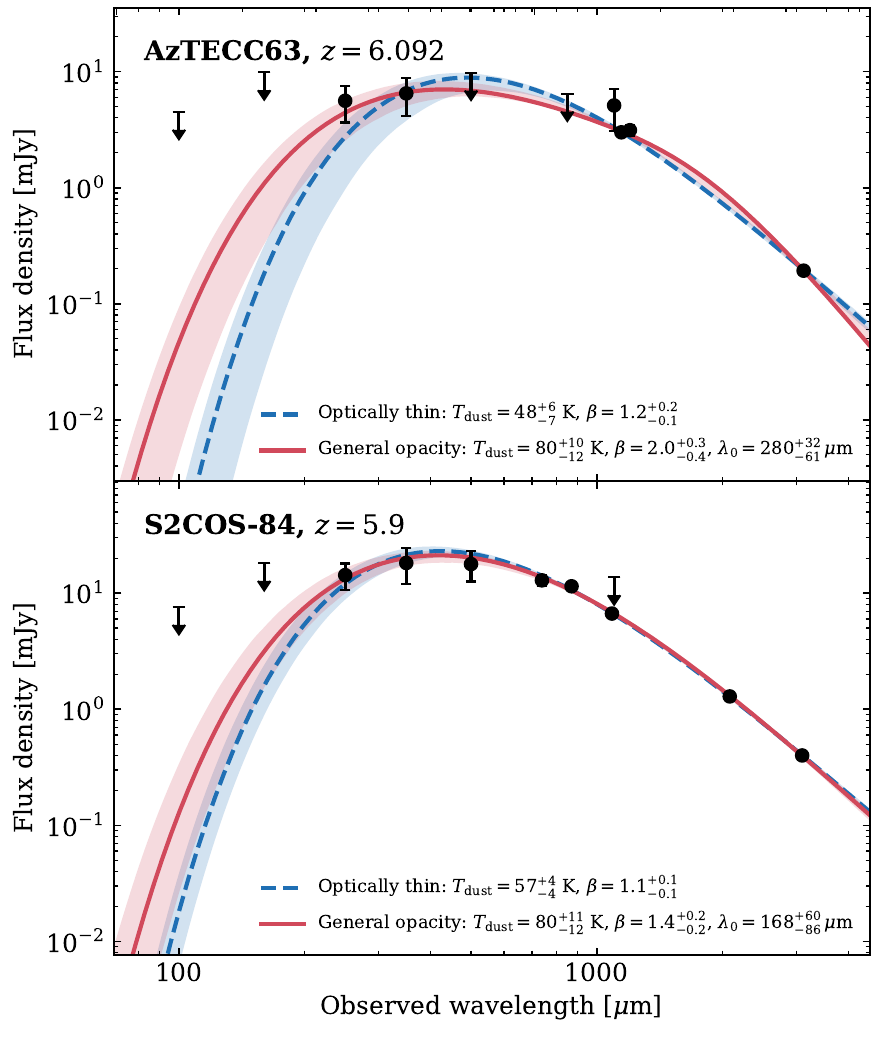}
\caption{FIR SEDs of AzTECC63 (top) and S2COS-84 (bottom) fitted with modified blackbody models including the effect of the CMB. The dashed blue and solid red curves show the best-fit optically thin and general-opacity models, respectively, and the shaded regions mark the 16th--84th percentile ranges of the posteriors. Arrows indicate $3\sigma$ upper limits. 
}
\label{fig:FIRsed}
\end{figure}

\end{appendix}

\end{document}